# Оптимальное легирование диодных прерывателей тока.

А. С. Кюрегян

Всероссийский Электротехнический институт им. В. И. Ленина, 111250, Москва, Россия
E-mail: ask@vei.ru

Получено аналитическое решение задачи об уменьшении потерь энергии $\Omega$ в диодных прерывателях тока на этапе восстановления блокирующей способности за счет оптимизации распределения легирующих примесей $N(x)$ по толщине структуры. Найдено распределение $N(x)$, близкое к оптимальному, позволяющее уменьшить $\Omega$ на 30-55% по сравнению с прерывателями стандартной конструкции с однородно легированными высокоомными слоями.

## 1. Введение.

Самыми эффективными диодными прерывателями тока являются так называемые дрейфовые диоды с резким восстановлением (ДДРВ) [1-4]. Аналитическая теория, позволяющая описать работу простейшего ДДРВ на основе $n^+$-$p$-$p^+$-структуры с однородно легированной высокоомной $p$-базой[1], была построена в работе автора [5]. Один из результатов этой работы состоит в том, что при прочих равных условиях улучшение электрических характеристик генераторов возможно только за счет увеличения энергии $\Omega$, рассеиваемой ДДРВ на этапе восстановления блокирующей способности, и соответствующего снижения коэффициента полезного действия (см. Рис. 5 в [5]). Эта энергия рассеивается главным образом омическим сопротивлением $r(l)$ той части $p$-базы $l < x < w$, которая свободна от неравновесной плазмы и расширяется со временем вплоть до начала резкого обрыва тока (см. Рис. 1). В работе [5] при расчете $\Omega$ использовалось простое феноменологическое соотношение между $r(l)$ и извлеченным из базы зарядом неравновесных дырок $Q(l)$, полученное путем анализа результатов численного моделирования работы ДДРВ в типичных режимах для случая однородно легированной $p$-базы. Разумеется, такой подход не позволяет ответить на вопрос о том, можно ли уменьшить энергию $\Omega$ путем модификации профиля легирования базы. В настоящей работе мы решим эту задачу, используя физически более содержательную модель процесса восстановления блокирующей способности высоковольтных диодов [7], описанную в разделе 2. Там же сформулирована вариационная задача оптимизации профиля легирования $N(x)$ высокоомной $n^+$-$p$-$p^+$-структуры. В разделе 3 проведена оптимизация параметров однородно легированной базы. В разделе 4 изучены основные свойства оптимального профиля $\tilde{N}(x)$, а в разделе 5 получено приближенное решение $N_1(x)$, близкое к оптимальному. Наконец, в Заключении изложены основные результаты работы.

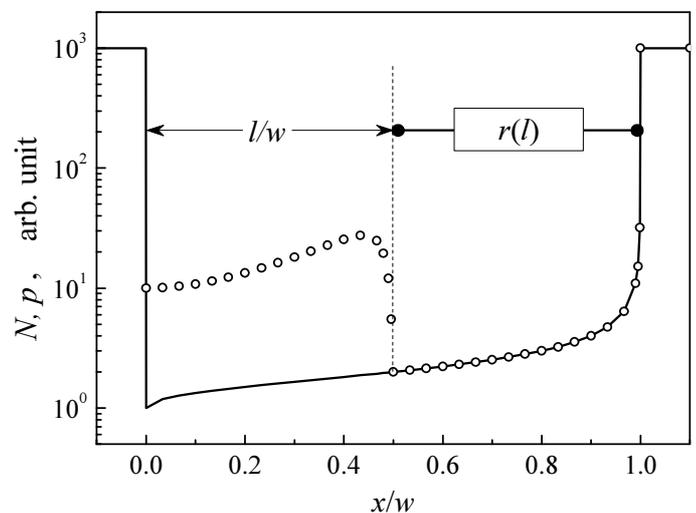

Рис. 1. . Схематические распределения легирующих примесей $N(x)$ (сплошная линия) и дырок $p(x)$ в $n^+$-$p$-$p^+$-структуре во время фазы высокой обратной проводимости (символы).

## 2. Модель процесса восстановления и формулировка вариационной задачи.

Рассмотрим $n^+ - p - p^+ -$ структуру, $p -$ база которой имеет толщину $w$ и неоднородно легирована акцепторами с концентрацией $N(x)$. По-сле окончания прямой накачки база заполнена квазинейтральной плазмой с концентрацией дырок $p_0(x) \gg N(x)$. Протекание тока обратной полярности с плотностью $j(t)$ приводит к восстановлению $p-p^+$–перехода в момент времени $t=0$ и образованию фронта, движущегося в сторону $n^+-p$–перехода. Фронт разделяет базу на две квазинейтральные области, в одной из которых $(0 < x < l)$ концентрация дырок $p(x) \approx p_0(x)$ и практически не изменяется со временем [8,9], а в другой $(l < x < w)$ неравновесная плазма отсутствует и $p(x) = N(x)$ (см. Рис.1). Если толщина фронта $\Delta l \sim q p_0(l) D_h / j$ ($D_h$ – коэффициент диффузии дырок, $q$ – элементарный заряд) много меньше $w$, то скорость фронта равна [7]

$$v_f \equiv -\frac{dl(t)}{dt} = \frac{\mu j(t)}{q \mu_h p[l(t)]}, \quad (1)$$

где $\mu = \mu_e \mu_h / (\mu_e + \mu_h)$, $\mu_{e,h}$ – низкополевые подвижности электронов и дырок. Эта стадия завершается в момент времени $t=T$, когда база полностью освобождается от неравновесной плазмы[2], то есть $l(T) = 0$ и

$$w = \frac{\mu}{q \mu_h} \int_0^T \frac{j(t)}{p_0[l(t)]} dt. \quad (2)$$

---
[1] Такой тип ДДРВ предпочтителен, если отношение подвижностей электронов и дырок достаточно велико, как, например, в 4H-SiC [6].

В силу неравенства $p_0(x) \gg N(x)$ основная мощность рассеивается в области $l < x < d$, где она равна

$$\frac{j^2}{q \mu_h} \int_l^w \frac{dx}{N(x) - j/q v_{sh}},$$

если ток смещения пренебрежимо мал, а дрейфовая скорость дырок зависит от напряженности поля $E$ по закону $v_h(E) = v_{sh} E / (E + E_{sh})$, $v_{sh} = \mu_h E_{sh}$ – насыщенная скорость дырок. При правильном согласовании параметров импульсов прямого и обратного токов с параметрами диода обратный ток достигает максимума $j_M$ при $t=T$. Нетрудно показать, что в этом случае за время $T$ в $p$–слое рассеивается энергия

$$\Omega = Q_0 \frac{v_{sh}}{\mu} \int_0^w \left[ \int_0^x \frac{f_j(l) f_p(l) dl}{q v_{sh} N(x)/j_M - f_j(l)} \right] \frac{dx}{w}, \quad (3)$$

где $f_p(x) = p_0(x)/\bar{p}_0$, $\bar{p}_0$ и $Q_0 = q \bar{p}_0 w$ – средняя концентрация и заряд дырок в базе при $t=0$. Функцию $f_j[l(t)] = j(t)/j_M$ и время $T$ легко найти для любых зависимостей $j(t)$, $f_p(l)$ из уравнений (1), (2).

Нам надо найти распределение примесей $N(x) = \tilde{N}(x)$, обеспечивающее минимум $\tilde{\Omega}$ функционала (3) при заданных значениях накопленного заряда $Q_0$, обрываемого тока $j_M$ и максимального падения напряжения

$$U_M = \int_0^w E(x) dx \quad (4)$$

на области пространственного заряда (ОПЗ), возникающей во время обрыва тока через диод и заполняющей всю базу при $U = U_M$. Разумеется, напряженность поля $E(x)$ должна быть положительной во всей ОПЗ и связана с $N(x)$ уравнением Пуассона

$$E'(x) = \frac{dE}{dx} = -\frac{q}{\varepsilon} N(x), \quad (5)$$

где $\varepsilon$ – диэлектрическая проницаемость полупроводника. Кроме того, $U_M$ не должно превышать напряжение $U_B$ пробоя диода[3], поэтому должно еще выполняться дополнительное неравенство

$$\int_0^w \alpha[E(x)] dx \leq 1, \quad (6)$$

где $\alpha(E)$ – эффективный коэффициент ударной ионизации. В настоящей работе изучен только практически наиболее важный случай, когда $U_M = U_B$ и (6) является равенством, которое, в частности, определяет величину максимальной напряженности поля $E(0) = E_B$ при пробое. Далее мы будем пренебрегать слабой зависимостью $E_B$ от профиля легирования, учет которой сильно усложняет вычисления, но приводит лишь к незначительным количественным изменениям результатов.

Результат оптимизации профиля легирования зависит от начального распределения плазмы $p_0(x)$ и формы импульса обратного тока. Кон-

---

[2] Заметим, что $n^+$-$p$-переход не успевает восстановиться до $t=T$ при больших отношениях $\mu_e/\mu_h$ [6].

[3] В противном случае энергия, рассеиваемая диодом во время обрыва тока, резко возрастает и может превысить $\Omega$.

кретные расчеты мы будем проводить, полагая для определенности, что

$$f_p(l) = a\frac{\operatorname{ch}(al/w)}{\operatorname{sh}(a)} \quad (7)$$

и $j = j_M$ или $j = j_M \sin(\pi t/2T)$. В первом случае $f_j = 1$, а во втором

$$f_j(l) = \sqrt{1 - \left[\int_0^l f_p(x)\frac{dx}{w}\right]^2} = \sqrt{1 - \left[\frac{\operatorname{sh}(al/w)}{\operatorname{sh}(a)}\right]^2}, \quad (8)$$

где $a$ – параметр, который зависит от длительности импульса прямой накачки ДДРВ и характеризует степень неоднородности начального распределения плазмы.

### 3. Однородно легированная база

Если концентрация акцепторов $N$ не зависит от $x$, то интегрирование (3) по частям приводит к формуле

$$\Omega = \Omega_0 = Q_0 \frac{v_{sh}}{\mu} \int_0^w \frac{f_j(x)f_p(x)(1-x/w)}{(2\omega-1)/i_M\omega^2 - f_j(x)}dx, \quad (9)$$

где использованы обозначения $\omega = w/w_0$, $i_M = j_M/qv_{sh}N_0$, $w_0 \equiv 2U_B/E_B$, $N_0 \equiv \varepsilon E_B^2/2qU_B$ и соотношение $qNw^2 = 2\varepsilon(E_B w - U_B)$, верное при $\omega \leq 1$.

В простейшем случае постоянного обратного тока $f_j = 1$ и подстановка (7) в (9) приводит после интегрирования к формуле

$$\Omega = \Delta\frac{\omega}{(2\omega-1)/i_M\omega^2 - 1}\frac{2}{a}\operatorname{th}\frac{a}{2}, \quad (10)$$

где $\Delta = Q_0 v_{sh} U_B/\mu E_B$. Обычно считают [2-5], что $w = w_0$. В этом случае

$$\Omega = \Omega_{00} = \Delta\frac{i_M}{1-i_M}\frac{2}{a}\operatorname{th}\frac{a}{2}. \quad (11)$$

Однако, как нетрудно убедиться, правая часть формулы (10) оказывается минимальной и равной

$$\Omega_{0m} = \Delta\frac{i_M\omega_0^2}{1-i_M\omega_0}\frac{3}{a}\operatorname{th}\frac{a}{2} \quad (12)$$

при оптимальных значениях $w_{0m} = w_0\omega_0$ и $N_m = N_0\eta_0$, где

$$\omega_0 = 2i_M^{-1}\left(1-\sqrt{1-3i_M/4}\right) \text{ и } \eta_0 = \omega_0^{-2}(2\omega_0-1) \quad (13)$$

Зависимости $\omega_0(i_M)$ и $\eta(i_M)$ приведены на Рис. 2. Для функции $f_j(l)$, определяемой формулой (8), оптимальное значение безразмерной толщины ба-зы $\bar{\omega}_0$ и соответствующую минимальную энергию $\bar{\Omega} = \Omega(\bar{\omega}_0)$, а также энергию $\Omega_0$, рассеиваемую диодом с $w = w_0$ и $N = N_0$, легко найти численными методами. Результаты расчетов, приведенные на Рис. 2, показывают, что оптимизация величин $N = const$ и $w$ приводит к снижению энергии потерь на стадии восстановления блокирующей способности, которое не зависит от $a$. Однако выигрыш не превосходит 16% при $j_M \ll 1$ и исчезает при $j_M \to 1$ (см. Рис. 2).

Кроме того оказалось, что при всех актуальных значениях параметра $a$ выполняются неравенства

$0 < \omega_{0m}/\bar{\omega}_0 - 1 < 0.012$ и $0 < \Omega(\omega_{0m})/\bar{\Omega} - 1 < 0.001$.

Отсюда следует важный для дальнейшего вывод: при $N = const$ минимизация функционала $\Omega[N(x)]$ после замены в знаменателе (9) функции $f_j(l)$ на единицу приводит к результату, который практически не отличается от точного.

### 4. Свойства точного решения вариационной задачи.

Используя уравнение Пуассона (5), формулу (3) можно переписать в виде

$$\Omega = -Q_0\frac{j_M}{\varepsilon\mu}\int_0^w\left[\int_0^x \frac{f_j(l)f_p(l)dl}{E'(x)+E'_s f_j(l)}\right]\frac{dx}{w}, \quad (14)$$

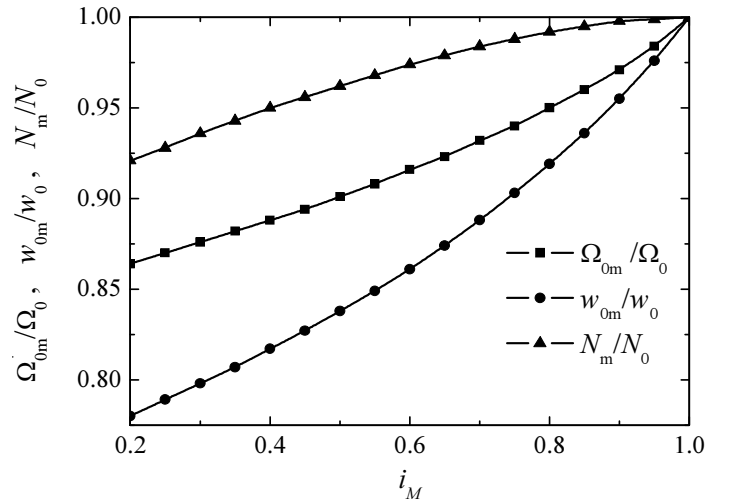

Рис. 2. Зависимости отношений $\Omega_{0m}/\Omega_0$ (квадраты), $w_{0m}/w_0$ (кружки) и $N_m/N_0$ от безразмерной плотности тока $i_M$. Линии - $j = j_M \sin(\pi t/2T)$, символы - $j = j_M$.

где $E'_s = j_M/\varepsilon v_{sh}$. Определение функции $E(x)$, которая обеспечивает минимум $\tilde{\Omega}$ функционала (14) при дополнительных условиях (4) и (6), является классической изопериметрической задачей вариационного исчисления [10]. Ее решение должно удовлетворять уравнению Эйлера

$$\frac{d}{dx}\frac{\partial F}{\partial E'} = \frac{\partial F}{\partial E} \quad (15)$$

для функции

$$F(x,w,E,E') = -\frac{1}{w}\int_0^x \frac{f_j(l)f_p(l)dl}{E'(x)+E'_s f_j(l)} - C_1 E + C_2 \alpha(E) \quad (16)$$

где $C_{1,2}$ - неопределенные множители Лагранжа. Подстановка (16) в (15) приводит к уравнению

$$\frac{d}{dx}\frac{\partial F}{\partial E'} = -C_1 + C_2 \alpha'_E, \quad (17)$$

первый интеграл которого имеет вид

$$\frac{\partial F}{\partial E'} = \frac{1}{w}\int_0^x \frac{f_j(l)f_p(l)dl}{\left[E'(x)+E'_s f_j(l)\right]^2} = -C_1 x + C_2 \int_0^x \alpha'_E dx + C, \quad (18)$$

где $\alpha'_E = d\alpha(E)/dE$. Граничные значения напряженности поля $E_B = E(0)$, $E_w = E(w)$ и положение границы $x = w$ заранее неизвестны. Поэтому [10] при $x = 0$ следует принять «естественное» граничное условие

$$\partial F/\partial E'\big|_{x=0} = 0, \quad (19)$$

которое выполняется автоматически, а при $x = w$ – условия трансверсальности

$$\partial F/\partial E'\big|_{x=w} = 0, \quad (20)$$

$$F\big|_{x=w} + \int_0^w \frac{\partial F}{\partial w}dx = E'(w)\frac{\partial F}{\partial E'}\bigg|_{x=w}. \quad (21)$$

Из (18)-(20) следует, что $C = 0$, $1/E'(w) = 0$ и $C_1 w = C_2 \int_0^w \alpha'_E dx$. Используя эти равенства и теорему о среднем, интегро-дифференциальное уравнение (18) можно переписать в виде

$$E'(x) = -\frac{j_M}{\varepsilon v_{sh}}\left[f_j(\hat{x}) + \lambda\sqrt{\frac{\varphi(x)}{f_\alpha(x) - x/w}}\right], \quad (22)$$

где

$$\varphi(x) = w^{-1}\int_0^x f_j f_p dl, \quad \lambda = \frac{\varepsilon v_{sh}}{j_M}(C_1 w)^{-1/2},$$

$$f_\alpha(x) = \int_0^x \alpha'_E dx \left(\int_0^w \alpha'_E dx\right)^{-1},$$

а неизвестная координата $\hat{x}$ зависит от $x$ и удовлетворяет неравенствам $0 \le \tilde{x} \le x$. Кроме того из (18) и (22) следует, что

$$E'\frac{\partial F}{\partial E'}\bigg|_{x\to w} \propto \sqrt{w-x}\bigg|_{x\to w} \to 0,$$

поэтому граничное условие (21) принимает вид

$$\left(\frac{\lambda j_m}{\varepsilon v_{sh}}\right)^{-2}\left\{\alpha\big[E(w)\big]\left[\int_0^w \alpha'_E dx\right]^{-1} - \frac{E(w)}{w}\right\} + $$
$$+ w^{-2}\int_0^w\left[\int_0^x \frac{f_j(l)f_p(l)dl}{E'(x)+E'_s f_j(l)}\right]dx = 0 \quad . \quad (23)$$

Нетрудно убедиться в том, что левая часть этого равенства, пропорциональная $\partial\Omega/\partial w$, меньше нуля при любой реалистичной зависимости $\alpha(E)$ и монотонно падающей (см. (22)) функции $E(x) \ge 0$. Поэтому энергия $\Omega$ уменьшается с ростом $w$, но не достигает своего минимума, так как раньше нарушается условие $E(w) \ge 0$ и решение (22) теряет физический смысл. Таким образом, вместо (23) в нашем случае следует использовать граничное условие[4]

$$E(w) = 0. \quad (24)$$

Соотношение (22) содержит еще две неизвестные функции $f_j[\hat{x}(x)]$, $f_\alpha(x)$. Однако оно позволяет выяснить основные свойства точного решения $E'(x) = q\tilde{N}(x)/\varepsilon$ интегрального уравнения (18), обеспечивающего минимум функционалов (3) и (14). Действительно, при $x \to 0$

$$f_j \approx 1 - \frac{1}{2}\left[f_p(0)\frac{x}{w}\right]^2, \quad \varphi \approx f_p(0)\frac{x}{w},$$

$$f_\alpha \approx \alpha'(E_B)x\left(\int_0^w \alpha'_E dx\right)^{-1},$$

поэтому

$$\tilde{N}(0) \approx N_0 i_M\left[1 + \lambda\sqrt{\frac{f_p(0)}{\sqrt{\alpha'(E_B)w\left(\int_0^w \alpha'_E dx\right)^{-1}} - 1}}\right].$$

---

[4] Это граничное условие заранее совершенно неочевидно. Например, в случае $N = const$, рассмотренном в разделе 3, $E(w) > 0$ при оптимальных значениях $w$ и $N$.

Вследствие очень резкой зависимости $\alpha(E)$ толщина слоя эффективного умножения много меньше $w$. За ее пределами существует область, где еще $f_j[\hat{x}(x)] \approx 1$, но уже $f_\alpha(x) \approx 1$ и поэтому

$$\tilde{N}(x) \approx N_0 i_M \left[1 + \lambda \sqrt{\frac{\varphi(x)}{1 - x/w}}\right]. \quad (26)$$

При дальнейшем росте $x$ функция $f_j(\hat{x})$ становится заметно меньше 1, но одновременно с этим второе слагаемое в правой части (22) быстро возрастает и становится много больше $j_M f_j(\hat{x})/qv_{sh}$. Поэтому вблизи правой границы области $0 < x < w$ также можно использовать решение (26).

### 5. Приближенное решение вариационной задачи.

Точное аналитическое решение интегрального уравнения (18), определяющего неизвестные функции $\hat{x}(x)$ и $f_\alpha(x)$, невозможно. Однако далее будет показано, что даже использование профиля легирования $N_1(x)$, близкого к оптимальному $\tilde{N}(x)$, позволяет существенно уменьшить энергию $\Omega$ по сравнению с $\bar{\Omega}$. Для определения функции $N_1(x)$ мы используем два упрощения.

Во-первых, предположим, что минимизация функционала $\Omega[N(x)]$ после замены в знаменателе (9) функции $f_j(l)$ на единицу приводит к достаточно точным результатам, как и в случае с однородно легированной базой (см. конец раздела 3). Тогда из (22) получается

$$N_1(x) \approx N_0 i_M \left[1 + \lambda_1 \sqrt{\frac{\varphi(x)}{f_\alpha(x) - x/w_1}}\right]. \quad (27)$$

Примеры профилей, рассчитанных по этой формуле, приведены на Рис. 3. Интегрирование (5) с использованием (27) дает распределение напряженности поля в ОПЗ при $U = U_B$

$$E(x) = E_B \left\{1 - i_M \omega_1 \left[\frac{x}{w_1} + \lambda_1 \int_0^x \sqrt{\frac{\varphi}{f_\alpha(x) - x/w_1}} \frac{dx}{w_1}\right]\right\}, (28)$$

где $\omega_1 = w_1/w_0$.

Во-вторых, учтем, что в тонком слое умножения, где $f_\alpha(x)$ заметно отличается от единицы, функция $N_1(x)$ почти постоянна и равна $N_1(0)$ (см. Рис. 3). В этой области

$$f_\alpha[E(x)] \approx 1 - \alpha(E)/\alpha(E_B),$$
$$E(x) \approx E_B \left[1 - N_1(0)x/N_0 w_0\right].$$

Поэтому, используя обычную зависимость $\alpha(E) = \alpha_i \exp(-E_i/E)$, легко получить для $f_\alpha(x)$ аппроксимацию

$$f_\alpha(x) = 1 - \exp\left(\xi \frac{\eta_1 \omega_1 x/w_1}{\eta_1 \omega_1 x/w_1 - 1}\right) + \frac{x}{w_1} \exp\left(\xi \frac{\eta_1 \omega_1}{\eta_1 \omega_1 - 1}\right), \quad (29)$$

$$\eta_1 = \frac{N_1(0)}{N_0} = i_M \left[1 + \lambda_1 \sqrt{\frac{f_p(0)}{\xi \eta_1 \omega_1 - 1}}\right], \quad (30)$$

где $\xi = E_i/E_B$. Второе слагаемое в (29), много меньшее 1, введено для того, чтобы функция $f_\alpha(x)$ имела правильные значения 0 и 1 на обеих границах базы. Далее мы будем использовать типичное для высоковольтных диодов значение $\xi = 7.8$. Подстановка (28) в (4) и (24) позволяет найти формулы для $\omega_1$ и $\lambda_1$:

$$\omega_1 = \frac{1 - \sqrt{1 - i_M(1 - \chi^2)}}{i_M(1 - \chi)}, \quad \lambda_1 = \frac{1 - i_M \omega_1}{i_M \omega_1 I_E}, \quad (31)$$

$$\chi = \frac{I_-^0}{I_-^1} - 1, \quad I_\pm^\nu = \int_0^1 y^\nu \sqrt{\phi(y)[f_\alpha(w_1 y) - y]^{\pm 1}} dy, (32)$$

где $y = x/w$, $\phi(x/w) = \varphi(x)$. Интегралы $I_\pm^\nu$ и, следовательно, нормированная толщина базы $\omega_1$, зависят от одного заранее неизвестного параметра $\gamma = \eta_1 \omega_1$, который является решением уравнения (30), представимого в виде

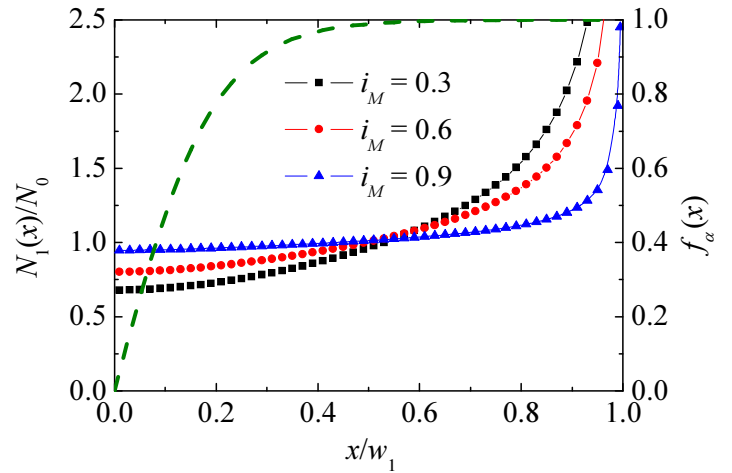

Рис. 3. «Оптимальные» профили легирования базы (символы) при $a = 0$ и различных $i_M$. Штриховая линия – функция $f_\alpha(x)$, рассчитанная по формуле (29).

$$\frac{\gamma - i_M \omega_1(\gamma)}{1 - i_M \omega_1(\gamma)} I_-^0(\gamma) = \sqrt{\frac{f_p(0)}{\xi\gamma - 1}} \qquad (33)$$

Таким образом, использованные нами два приближения позволили свести задачу оптимизации профиля легирования к простому численному решению уравнения (33), в котором функция $\omega_1(\gamma)$ определяется формулами (31), (32). Входящая в (32) функция $\phi(y)$ равна

$$\phi(y) = \operatorname{sh} ay / \operatorname{sh} a \qquad (34)$$

при $j(t) = j_M$ и $\phi(y) = \Phi(\operatorname{sh} ay / \operatorname{sh} a)$,

$$\Phi(z) = \frac{1}{2}\left(z\sqrt{1-z^2} + \arcsin z\right) \qquad (35)$$

при $j(t) = j_M \sin(\pi t / 2T)$. Зависимости $\omega_1(i_M, a)$, $\lambda_1(i_M, a)$ и $\eta_1(i_M, a)$, рассчитанные для этих двух случаев, приведены на Рис. 4, 5. Для энергии $\Omega$, соответствующей профилю легирования (27), в первом случае получается формула

$$\Omega = \Omega_{10} = \Delta \frac{i_M \omega_1^2}{1 - i_M \omega_1} 2 I_-^0 I_+^0 \qquad (36)$$

Во втором случае энергию $\Omega = \Omega_1$ можно получить только путем численного интегрирования (3) с использованием формул (27), (31)-(33) и (35). Результаты расчетов $\Omega_{10}(i_M, a)$ и $\Omega_1(i_M, a)$ приведены на Рис. 6.

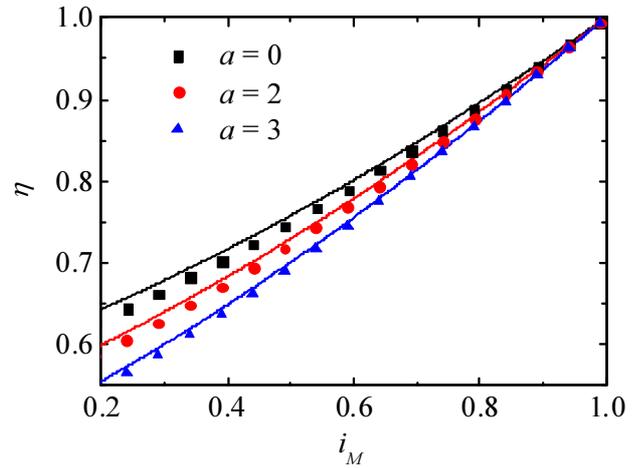

Рис. 5. Зависимости «оптимальных» значений нормированной граничной концентрации $\eta_1 = N_1(0)/N_0$ от нормированной плотности тока $i_M$ при различных $a$. Линии - $j = j_M \sin(\pi t / 2T)$, символы - $j = j_M$.

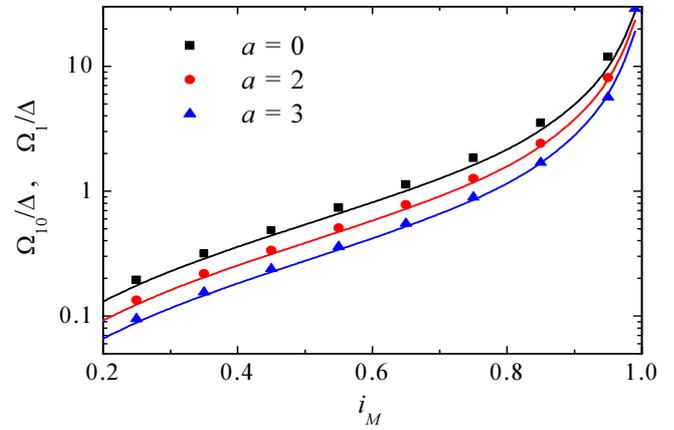

Рис. 6. Зависимости энергий $\Omega_1$ (линии, $j = j_M \sin(\pi t / 2T)$) и $\Omega_{10}$ (символы, $j = j_M$) от нормированной плотности тока $i_M$ при различных $a$.

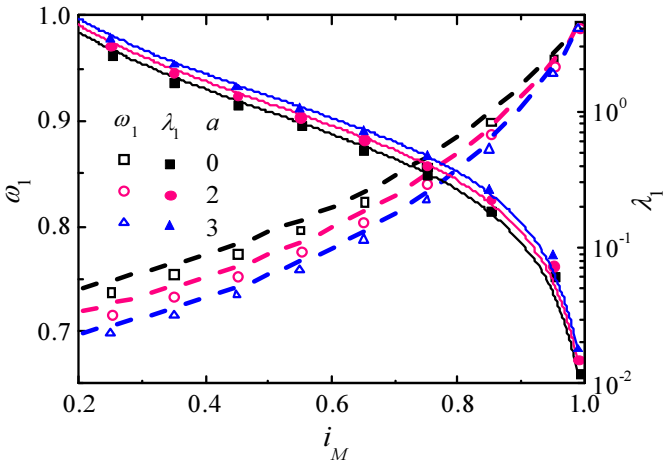

Рис. 4. Зависимости «оптимальных» значений нормированной толщины базы $\omega_1$ (светлые символы и штриховые линии) и параметра $\lambda_1$ (темные символы и сплошные линии) от нормированной плотности тока $i_M$ при различных $a$. Линии - $j = j_M \sin(\pi t / 2T)$, символы - $j = j_M$.

## 6. Заключение

Изложенная выше теория и основанные на ней результаты численных расчетов позволяют сделать следующие выводы.

1) Во всех рассмотренных случаях оптимальные толщина базы $w$ и концентрация примесей $N(x)$ стремятся к $w_0$ и $N_0$ соответственно при $j_M \to q v_{sh} N_0$ (см. Рис. 2-5). Поэтому оптимизация профиля легирования имеет смысл только при плотностях тока, заметно меньших $q v_{sh} N_0$.

2) Именно такой режим работы ДДРВ представляется целесообразным, поскольку энергия $\Omega$, рассеиваемая ДДРВ на этапе восстановления блокирующей способности, быстро увеличивает-ся с ростом обрываемой плотности тока $j_M$ (см. Рис. 6), хотя и не может стремиться к бесконечности при $j_M \to qv_{sh}N_0$, как это следует из формул (11), (12), (36). Причина этот «противоречия» состоит в том, что наша теория не учитывает неизбежное нарушение нейтральности и быстрый рост тока смещения в области $x > l$ при $j_M \to qv_sN_0$.

3) Результаты оптимизации распределения примесей в базе ДДРВ очень слабо зависят от формы импульса обратного тока $j(t)$ (Рис. 4-6). Изменения параметра $\xi$ в разумных пределах (на $\pm 10\%$) также практически не влияют на результаты. Это лишний раз указывает на то, что профиль легирования (27) очень близок к оптимальному.

4) Оптимизация толщины однородно легированной базы позволяет уменьшить энергию $\Omega$, но лишь на 5-12% при актуальных плотностях обрываемого тока $j_M = (0.8 - 0.4)qv_{sh}N_0$ (см. Рис. 2). В то же время данные, приведенные на Рис. 7, показывают, что использование профиля легирования (27), близкого к оптимальному, обеспечивает выигрыш 30-55% при тех же значениях $j_M$.

5) Энергия $\Omega$ уменьшается с ростом параметра $a$, характеризующего степень неоднородности начального распределения плазмы $p_0(x)$.

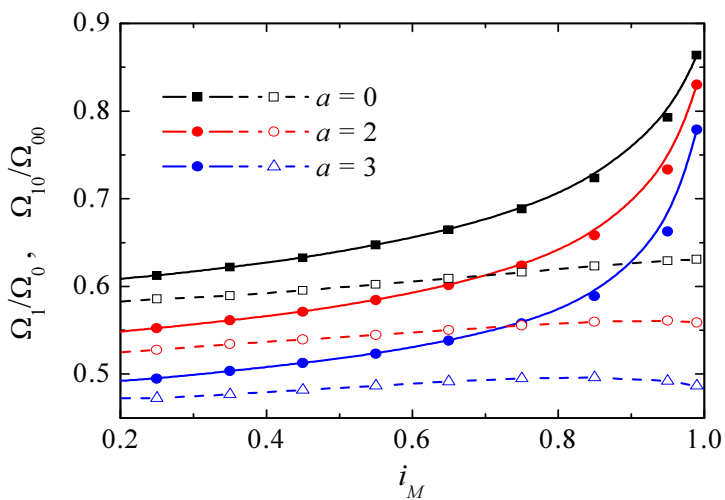

Рис. 7. Зависимости отношений энергий, рассеиваемых оптимизированным ДДРВ ($\Omega_1$ и $\Omega_{10}$), к энергиям, рассеиваемым стандартным ДДРВ ($\Omega_0$ и $\Omega_{00}$), от нормированной плотности тока $i_M$ при различных $a$. Штриховые линии и светлые символы - $j = j_M$. Сплошные линии и темные символы - $j = j_M \sin(\pi t/2T)$.

Причина этого состоит в том, что с ростом $a$ уменьшается длительность заключительной части процесса восстановления, во время которой сопротивление $r(l)$ и, следовательно, рассеиваемая мощность велики.

В заключение отметим, что аналогичную, но более громоздкую процедуру можно использовать для оптимизации профиля легирования ДДРВ на основе $n^+$-$n$-$p$-$p^+$-структур. Такие структуры необходимо использовать, если отношение $\mu_e/\mu_h$ не очень велико (как, например, в Si). В обоих высокоомных слоях распределения доноров и акцепторов должны описываться формулой (27), а соотношение между их толщинами должно обеспечивать главное условие эффективной работы ДДРВ [2,5,11] – одновременное достижение плоскости $n$-$p$-перехода обоими фронтами плазменной области.